\documentclass[prd,email,preprint,showkeys,preprintnumbers,amsmath,amssymb,nofootinbib]{revtex4}
\usepackage{graphicx}
\usepackage{amssymb}
\usepackage{epsfig}
\usepackage{amssymb,amsmath}
\usepackage{eufrak}
\usepackage{euscript}
\usepackage{latexsym}
\usepackage[latin1]{inputenc}
\usepackage{pstricks}

\newcommand{\be}{\begin{equation}}
\newcommand{\ee}{\end{equation}}
\newcommand{\ba}{\begin{eqnarray}}
\newcommand{\ea}{\end{eqnarray}}
\newcommand{\p}{\partial}

\skip\footins 20pt plus4pt minus4pt
\begin{document}

\title{\LARGE\textbf{Self-quartic interaction for a scalar field in an extended DFR noncommutative spacetime}}

\author{Everton M. C. Abreu$^{a,b}$}
\email{evertonabreu@ufrrj.br}
\author{M. J. Neves$^a$}
\email{mariojr@ufrrj.br}

\affiliation{${}^{a}$Grupo de F\' isica Te\'orica e Matem\'atica F\' isica, Departamento de F\'{\i}sica,
Universidade Federal Rural do Rio de Janeiro\\
BR 465-07, 23890-971, Serop\'edica, Rio de Janeiro, Brazil\\
${}^{b}$Departamento de F\'{\i}sica, ICE, Universidade Federal de Juiz de Fora,\\
36036-330, Juiz de Fora, MG, Brazil\\
\bigskip
\today\\}




\begin{abstract}


{\noindent The framework Doplicher-Fredenhagen-Roberts (DFR) of a noncommutative (NC) space-time
is considered as a alternative approach to study the NC space-time of the early Universe. In this formalism,
the parameter of noncommutative $\theta^{\mu\nu}$ is promoted to a coordinate of the space-time, and consequently,
we are describing a field theory in a space-time with extra-dimension. Consequently, there is a canonical momentum
associated to this new coordinate in which the effects of a new physics can emerge in the
propagation of the fields along the extra-dimension. The Fourier space of this framework is automatically
extended by the addition of new momenta components. The main concept that we would like to emphasize from the
outset is that the formalism demonstrated here will not be constructed introducing a NC parameter in the system,
as usual. It will be generated naturally from an already NC space. When the components of the new momentum are zero,
the DFR approach is reduced to the usual NC case, in which $\theta^{\mu\nu}$ is a antisymmetric constant matrix.
We study a scalar field action with self-quartic interaction $\phi^{4}\star$ defined in the DFR NC spacetime,
obtaining the Feynman rules in the Fourier space for the scalar propagator and vertex of the model.
With these rules we are able to build out the radiative corrections to one loop order for the model propagator.
The influence of the NC scale, as well as the propagation of the field in the extra-dimension,
are analyzed in the ultraviolet divergences scenario. We investigate the actual possibility if this $\theta^{\mu\nu}$
conjugate momentum has the property of healing the mixing
IR/UV divergences that emerges in this recently new NC spacetime quantum field theory. } \\ \vspace{0.7cm}

\end{abstract}

\maketitle

\pagestyle{myheadings}
\markright{Self-quartic interaction for a scalar field in an extended DFR....}

\section{Introduction}
\renewcommand{\theequation}{1.\arabic{equation}}
\setcounter{equation}{0}

The understanding of Physics of the early Universe is one the great puzzles of the current science.
The idea of combining the fundamental physical constants like $\hbar, c$ and $G$ to obtain a length scale was given
by M. Planck in $1900$
\begin{eqnarray}\label{lp}
\ell_{P}=\sqrt{\frac{\hbar G}{c^{3}}}\sim 1.6 \times 10^{-33} \; \mbox{m} \; ,
\end{eqnarray}
where $\ell_{P}$ is the Planck's length. This relation shows that there is a mixing of quantum phenomena $(\hbar)$,
and gravitational world $(G)$ at this length scale.  Indeed, it is believed that effects of a
quantum gravity theory must emerge next to the Planck's scale. The introduction of a length scale in a physics theory
was only forty seven years after Planck's idea by constructing a noncommutative (NC) spacetime.
The motivation to introduce such scale was the need to tame the ultraviolet divergences
in quantum field theory (QFT). The first published work concerning a NC concept of space-time
was carried out in $1947$ by Snyder in his seminal paper \cite{snyder47}.

The central NC idea is that the space-time coordinates $x^{\mu}\;(\mu=0,1,2,3)$
are promoted to operators in order to satisfy the basic commutation relation
\begin{eqnarray} \label{xmuxnu}
\left[\,\hat{x}^\mu\,,\,\hat{x}^\nu\,\right]\,=\,i\,\ell\theta^{\mu\nu}\,\,,
\end{eqnarray}
where $\theta^{\mu\nu}$ is an antisymmetric constant matrix, and $\ell$ is a length scale.
The alternative would be to construct a discrete space-time with a NC algebra.
Consequently, the coordinates operators are quantum observable that satisfy the uncertainty relation
\begin{eqnarray}\label{uncertainxmu}
\Delta \hat{x}^{\mu} \Delta \hat{x}^{\nu} \simeq \ell \theta^{\mu\nu} \; ,
\end{eqnarray}
it leads to the interpretation that noncommutativity (NCY) of spacetime must emerge in
a fundamental length scale $\ell$, {\it i.e.} the {\it Planck's scale}, for example.

However, Yang \cite{yang47}, a little time later, demonstrated that Snyder's hopes in cutting off the infinities in QFT
were not obtained by NCY.   This fact doomed Snyder's NC theory to years of ostracism.
After the important result that the algebra obtained with a string theory embedded in a magnetic
background is NC, a new perspective concerning NCY was rekindle \cite{seibergwitten99}.
Nowadays, the NC quantum field theory (NCQFT) is one of the most investigated subjects about the description
of underlying physics at a fundamental length scale of quantum gravity \cite{QG}.


The most popular NCY formalism consider $\theta^{\mu\nu}$
as a constant matrix but different from (\ref{xmuxnu}), this formalism is called canonical NCY and is given by 
\begin{eqnarray}\label{1.4}
[x^{\mu}, x^{\nu}]=i\,\theta^{\mu\nu}\,\,.
\end{eqnarray}

Although it maintains the translational invariance,
the Lorentz symmetry is not preserved \cite{Szabo03}.   For example, in the case of the hydrogen atom, it breaks the rotational symmetry of the model, which removes the degeneracy of the energy levels \cite{Chaichian}.
To heal this disease a recent approach was introduced
by Doplicher, Fredenhagen and Roberts (DFR) \cite{DFR}. It considers $\theta^{\mu\nu}$ as an ordinary
coordinate of the system in which the Lorentz symmetry is preserved.
Recently, it has emerged the idea \cite{Morita} of constructing an extension of this so-called DFR spacetime
introducing the conjugate canonical momenta associated with
$\theta^{\mu\nu}$ \cite{Amorim1} (for a review the reader can see \cite{amo}).
This extended NC spacetime has ten dimensions:
four relative to Minkowski spacetime and six relative to $\theta$-space.
This new framework is characterized by a field theory constructed in a space-time with extra-dimensions $(4+6)$,
and which does not need necessarily the presence of a length scale $\ell$ localized into the six dimensions
of the $\theta$-space, where, from (\ref{1.4}) we can see that $\theta^{\mu\nu}$ now has dimension of length-square.
Besides the Lorentz invariance was recovered, and obviously we hope that causality aspects in
QFT in this $\left(x+\theta\right)$ space-time must be preserved too \cite{EMCAbreuMJNeves2012}.

By following this concept, the algebraic structure of this DFR-extended (which will be called DFR$^*$ from now on) NC phase-space is enlarged by
introducing the momenta operators associated to the coordinates $x^{\mu}$
and to the new coordinates $\theta^{\mu\nu}$. We have the usual canonical momenta $p_{\mu}$
associated to coordinates $x^{\mu}$, and for simplicity, we call $k_{\mu\nu}$ the
antisymmetrical canonical momentum associated to the new coordinate $\theta^{\mu\nu}$.
These objects are promoted to operators in an also extended Hilbert space ${\cal H}$ \cite{Amorim1,Amorim4,Amorim5,Amorim2}.
All the corresponding operators belong to the same algebra and have the same hierarchical level.   Recently it was demonstrated \cite{ammo} that the canonical momentum $k_{\mu\nu}$ is in fact connected to the Lorentz invariance of the system.   Hence, it is possible to construct a filed theory with this extended phase-space \cite{letter}.

The addition of the canonical momenta $k_{\mu\nu}$ is important if we
are interested in the investigation of the propagation of fields along the $\theta$-space. 
Our objective here is to disclose and analyze new physical aspects that
can emerge from the propagation of fields through the $\theta$-direction.


In this paper we obtain the one loop corrections to the scalar propagator NC $\phi^{4}\star$
defined in the DFR$^*$ space. We propose a NC action with a self quartic interaction $\phi^{4}\star$,
and the Feynman rules, necessary to perturbation theory are obtained.
Another target here is to study the primitive divergences of this model
and to calculate the one loop correction to the scalar propagator
for the cases in which $k_{\mu\nu}=0$, which is a NC toy model, as we will see, 
and $k_{\mu\nu}\neq 0$.
The existence of the NC length scale can bring something new to the mass
correction in the propagator due to the perturbation theory.
We will also investigate the influence of a extended Fourier space $(p_{\mu},k_{\mu\nu})$
in the primitive divergence of the propagator.

The paper is organized as: the next section is dedicated (for self-containment of this
work) to a review of the basics of NC quantum field theory DFR$^*$ framework, namely, the
DFR$^*$ algebra. In section 3 we write the action of scalar field with $\phi^{4}\star$ interaction,
and consequently, the Feynman rules in the Fourier space.
Section 4 is dedicated to the primitive divergences of model. We use the Feynman rules to obtain the expressions of the self-energy
at the one loop approximation. We divide it in two subsections, where in the first one we have the case of the self-energy with
$k^{\mu\nu}=0$ (toy model). The second subsection describes the case when $k_{\mu\nu}\neq 0$, in which the self-energy is calculated to investigate
the consequence of this momentum in the ultraviolet divergences.
To finish, we discussed the results obtained and we depict the final remarks and conclusions.





\section{The DFR$^*$ NC spacetime}
\renewcommand{\theequation}{2.\arabic{equation}}
\setcounter{equation}{0}
\subsection{The algebra}

In this section, we will review the main steps published in \cite{Amorim1,Amorim4,Amorim5,Amorim2}.
Namely, we will revisit the basics of the quantum field theory defined in the DFR$^*$ space.
The space-time coordinates $x^{\mu}=(t,{\bf x})$ do not commute itself
satisfying the commutation relation (\ref{xmuxnu}). The parameters $\theta^{\mu\nu}$ are promoted
to coordinates of this spacetime, which has $D=10$ and it has six independents spatial coordinates
associated to $\theta^{\mu\nu}$. Consequently, the parameters $\theta^{\mu\nu}$ are promoted
to operators $\hat{\theta}^{\mu\nu}$ in the commutation relation. 

Let us begin with the standard DFR algebra
\cite{DFR} involving only the position operators
\begin{eqnarray}\label{algebraDFRxtheta}
\left[\hat{x}^{\mu},\hat{x}^{\nu}\right] = i\hat{\theta}^{\mu\nu}
\hspace{0.2cm} , \hspace{0.2cm}
\left[\hat{x}^{\mu},\hat{\theta}^{\nu\alpha}\right] = 0
\hspace{0.3cm} \mbox{and} \hspace{0.3cm}
\left[\hat{\theta}^{\mu\nu},\hat{\theta}^{\alpha\beta}\right] = 0 \; .
\end{eqnarray}
\noindent notice that we cannot say that $\theta^{\mu\nu}$ are any kind of position operator because, although it is a coordinate in the DFR$^*$ space, it does not mean that it provide any kind of localization, like $x^{\mu}$ do in standard commutative spacetime.   It can be shown that this well known standard DFR space is in fact incomplete.  As a matter of fact, in \cite{ammo}, one of us showed that the existence of the canonical momentum $k_{\mu\nu}$ is intrinsically connected with $\theta^{\mu\nu}$ and, consequently, with the Lorentz invariance.  That is the reason we have called a DFR$^*$ system (that will be analyzed here) where the momentum associated to $\theta^{\mu\nu}$ is zero, a toy model.  In other words, if $\theta^{\mu\nu}$ are coordinates, there exists $k_{\mu\nu}$.

Hence,  the canonical conjugate momenta operator $\hat{k}_{\mu\nu}$
associated with the operator $\hat{\theta}^{\mu\nu}$ must satisfy the commutation relation
\begin{equation}\label{thetapicomm}
\left[\,\hat{\theta}^{\mu\nu},\hat{k}_{\rho\sigma}\, \right] = i \delta^{\mu\nu}_{\,\,\,\,\,\,\,\rho\sigma} \; ,
\end{equation}
where $\delta^{\mu\nu}_{\,\,\,\,\,\;\rho\sigma}=\delta^{\mu}_{\;\;\rho}\delta^{\nu}_{\;\;\sigma}
-\delta^{\mu}_{\;\;\sigma}\delta^{\nu}_{\;\;\rho}$. In order to obtain consistency we can write that \cite{Amorim1}
\begin{eqnarray}\label{algebraDFRpk}
\left[\hat{x}^{\mu},\hat{p}^{\nu} \right] = i\eta^{\mu\nu}
\hspace{0.2cm} , \hspace{0.2cm}
\left[\hat{p}^{\mu},\hat{p}^{\nu} \right] = 0
\hspace{0.2cm} , \hspace{0.2cm}
\left[\hat{\theta}^{\mu\nu},\hat{p}^{\rho}\right] = 0 \; ,
\nonumber \\
\left[\hat{p}^{\mu},\hat{k}^{\nu\rho}\right] = 0
\hspace{0.2cm} , \hspace{0.2cm}
[\hat{x}^{\mu},\hat{k}_{\nu\rho}]=-{i\over 2}\delta_{\nu\rho}^{\;\;\;\;\mu\sigma}\hat{p}_{\sigma} \; ,
\end{eqnarray}
and this completes the DFR$^*$ algebra. It is possible to verify that the whole set of commutation
relations listed above is indeed consistent with all possible Jacobi identities and the CCR algebras
\cite{EMCAbreuMJNeves2012}. The $\theta^{\mu\nu}$ coordinates
are constrained by the quantum conditions
\begin{eqnarray}\label{condtheta}
\theta_{\mu\nu}\theta^{\mu\nu}=0
\hspace{0.5cm} \mbox{and} \hspace{0.5cm}
\frac{1}{4}\star\!\theta_{\mu\nu}\theta^{\mu\nu}=\ell_{P}^{\;4} \; ,
\end{eqnarray}
where $\star\theta_{\mu\nu}=\varepsilon_{\mu\nu\rho\sigma}\theta^{\rho\sigma}$
and $\lambda_{P}$ is the Planck length. Here we have adopted that
$(\hbar=c=\ell=1)$ and consequently the coordinates $\theta^{\mu\nu}$ have dimension
of length to the squared. The uncertainty principle (\ref{xmuxnu}) is altered by
\begin{eqnarray}\label{uncertainxmu}
\Delta \hat{x}^{\mu} \Delta \hat{x}^{\nu} \simeq \langle \hat{\theta}^{\mu\nu}\rangle \; ,
\end{eqnarray}
in which the expected value of the operator $\hat{\theta}$ is related to the particles fluctuation position.

The last commutation relation of (\ref{algebraDFRpk}) suggests that the shifted coordinate
operator \cite{Chaichian,Gamboa,Kokado,Kijanka,Calmet1,Calmet2}
\begin{equation}\label{X}
\hat{X}^{\mu} = \hat{x}^{\mu}\,+\,{i\over 2}\hat{\theta}^{\mu\nu}\hat{p}_{\nu}\,\,,
\end{equation}
commutes with $\hat{k}_{\mu\nu}$.  The relation (\ref{X}) is also known as Bopp shift in the literature.
The commutation relation (\ref{algebraDFRpk})
also commutes with $\hat{\theta}^{\mu\nu}$ and $\hat{X}^{\mu}$, and satisfies a non trivial
commutation relation with $\hat{p}^{\,\mu}$ dependent objects, which could be derived from
\begin{equation}\label{Xpcomm}
[\hat{X}^{\mu},\hat{p}^{\nu}]=i\eta^{\mu\nu}
\hspace{0.6cm} , \hspace{0.6cm}
[\hat{X}^{\mu},\hat{X}^{\nu}]=0\,\,
\end{equation}
and one can note that the property $\hat{p}_{\mu} \hat{X}^{\mu}=\hat{p}_{\mu} \hat{x}^{\mu}$ can be easily verified.
Hence, we see from these both equations that the shifted coordinated operator (\ref{X}) allows us to recover
the commutativity property. The shifted coordinate operator $\hat{X}^{\mu}$ plays a fundamental role in NC
quantum mechanics defined in the $\left(x+\theta\right)$-space, since it is possible to form a basis with its eigenvalues.
This possibility is forbidden for the usual coordinate operator $\hat{x}^{\mu}$ since its components satisfy
nontrivial commutation relations among themselves (\ref{algebraDFRxtheta}). So,
differently from $\hat{x}^{\mu}$, we can say that $\hat{X}^{\mu}$ forms a basis in
Hilbert space.
The generator of Lorentz group is
\begin{eqnarray}\label{Mmunu}
\hat{M}_{\mu\nu}=\,\hat{X}_{\mu}\hat{p}_{\nu}\,-\,\hat{X}_{\nu}\hat{p}_{\mu}
+\hat{\theta}_{\nu\rho}\hat{k}^{\rho}_{\;\mu}-\hat{\theta}_{\mu\rho}\hat{k}^{\rho}_{\;\nu} \; ,
\end{eqnarray}
and from (\ref{algebraDFRpk}) we can write the generators
for translations as $\hat{p}_{\mu} \rightarrow - i \partial_{\mu}\,\,$.
With these ingredients it is easy to construct the commutation relations
\begin{eqnarray}\label{algebraMP}
\left[ \hat{p}_\mu , \hat{p}_\nu \right] &=& 0
\hspace{0.2cm} , \nonumber \\
\left[ \hat{M}_{\mu\nu},\hat{p}_{\rho} \right] &=& \,i\,\big(\eta_{\mu\rho}\,\hat{p}_\nu
-\eta_{\mu\nu}\,\hat{p}_\rho\big) \; ,
\hspace{0.1cm} \nonumber \\
\left[\hat{M}_{\mu\nu} ,\hat{M}_{\rho\sigma} \right] &=& i\left(\eta_{\mu\sigma}\hat{M}_{\rho\nu}
-\eta_{\nu\sigma}\hat{M}_{\rho\mu}-\eta_{\mu\rho}\hat{M}_{\sigma\nu}+\eta_{\nu\rho}\hat{M}_{\sigma\mu}\right) \; ,
\end{eqnarray}
it closes the appropriated algebra, and we can say that $\hat{p}_\mu$ and $\hat{M}_{\mu\nu}$
are the generators of the DFR$^*$ algebra.
Analyzing the Lorentz symmetry in NCQM following the lines above,
we can introduce an appropriate theory, for instance, given by a scalar action.
It is well known, however, that elementary particles are classified according
to the eigenvalues of the Casimir operators of the inhomogeneous Lorentz group.
Hence, let us extend this approach to the Poincar\'e group ${\cal P}$.
Considering the operators presented here, we can in principle consider that
\begin{eqnarray}\label{ElementoG}
\hat{G}={1\over2}\omega_{\mu\nu}\hat{M}^{\mu\nu}
-a^\mu\hat{p}_{\mu}
+{1\over2}b_{\mu\nu}\hat{k}^{\mu\nu} \; ,
\end{eqnarray}
is the generator of some group  ${\cal P}'$, which has the Poincar\'e group as a subgroup.
By defining the dynamical transformation of an arbitrary operator $\hat{A}$
in ${\cal H}$ in such a way that $\delta \hat{A}\,=\,i\,[\hat{A},\hat{G}]$
we arrive at the set of transformations,
\begin{eqnarray}\label{transf}
\delta \hat{x}^{\mu}&=&\omega ^\mu_{\,\,\,\,\nu}\hat{x}^{\nu}+a^\mu\nonumber\\
\delta\hat{p}_\mu&=&\omega _\mu^{\,\,\,\,\nu}\hat{p}_\nu\nonumber\\
\delta\hat{\theta}^{\mu\nu}&=&\omega ^\mu_{\,\,\,\,\rho}\hat{\theta}^{\rho\nu}
+ \omega ^\nu_{\,\,\,\,\rho}\hat{\theta}^{\mu\rho}+b^{\mu\nu}\nonumber\\
\delta\hat{k}_{\mu\nu}&=&\omega _\mu^{\,\,\,\,\rho}\hat{k}_{\rho\nu}
+ \omega _\nu^{\,\,\,\,\rho}\hat{k}_{\mu\rho}\nonumber\\
\delta \hat{M}_1^{\mu\nu}&=&\omega ^\mu_{\,\,\,\,\rho}\hat{M}_1^{\rho\nu}
+ \omega ^\nu_{\,\,\,\,\rho}\hat{M}_1^{\mu\rho}+a^\mu\hat{p}^\nu-a^\nu\hat{p}^\mu\nonumber\\
\delta \hat{M}_2^{\mu\nu}&=&\omega ^\mu_{\,\,\,\,\rho}\hat{M}_2^{\rho\nu}
+ \omega ^\nu_{\,\,\,\,\rho}\hat{M}_{2}^{\mu\rho}+b^{\mu\rho}\hat{k}_\rho^{\,\,\,\,\nu}
+ b^{\nu\rho}\hat{k}_{\,\,\,\rho}^{\mu}\nonumber\\
\delta \hat{x}^{\mu}&=&\omega ^\mu_{\,\,\,\,\nu}\hat{x}^{\nu}+a^\mu+{1\over2}b^{\mu\nu}\hat{p}_{\nu} \; .
\end{eqnarray}

One can observe that there is an unexpected term in the last equation of (\ref{transf}).
This is a consequence of the coordinate operator in (\ref{X}), which is a nonlinear combination
of operators that act on different Hilbert spaces.

The action of ${\cal P}'$ over the Hilbert space operators is in some sense equal to the action of the
Poincar\'e group with an additional translation operation on the ($\hat{\theta}^{\mu\nu}$) sector.
Its generators, all of them, close in an algebra under commutation. Hence, ${\cal P}'$ is a well
defined group of transformations. As a matter of fact, the commutation of two transformations closes in the algebra
\begin{equation}\label{algebray}
[\delta_2,\delta_1]\,\hat{y}=\delta_3\,\hat{y} \; ,
\end{equation}
where ${\mathbf y}$ represents any one of the operators appearing in (\ref{transf}).
The parameters composition rule is given by
\begin{eqnarray}\label{DecompOmega}
\omega^\mu_{3\,\,\nu}&=&\omega^\mu_{1\,\,\,\,\alpha}\omega^\alpha_{2\,\,\,\,\nu}-\omega^\mu_{2\,\,\,\,\alpha}\omega^\alpha_{1\,\,\,\,\nu}\nonumber\\
a_3^\mu&=&\omega^\mu_{1\,\,\,\nu}a_2^\nu-\omega^\mu_{2\,\,\,\nu}a_1^\nu \nonumber  \\
b_3^{\mu\nu}&=&\omega^\mu_{1\,\,\,\rho}b_2^{\rho\nu}-\omega^\mu_{2\,\,\,\rho}b_1^{\rho\nu}-\omega^\nu_{1\,\,\,\rho}b_2^{\rho\mu}+
\omega^\nu_{2\,\,\,\rho}b_{1}^{\rho\mu} \,\,.
\end{eqnarray}

\subsection{DFR$^*$ quantum mechanics and field theory}

To sum up, the framework showed above demonstrated that in NCQM, the physical coordinates
do not commute and the respective eigenvectors cannot be used to form a basis
in ${\cal H}={\cal H}_1 \oplus {\cal H}_2$ \cite{Amorim4}.  This can be accomplished
using the Bopp shift defined in
(\ref{X}) with (\ref{Xpcomm}) as consequence.
So, we can introduce a coordinate basis
$|X^{\prime}, \theta^{\prime} \rangle = |X^{\prime}\rangle \otimes |\theta^{\prime}\rangle$
and $|p^{\prime}, k^{\prime} \rangle = |p^{\prime}\rangle \otimes |k^{\prime}\rangle$,
in such a way that
\begin{eqnarray}\label{eqeigenvaluesXtheta}
\hat{X}^{\mu}|X^{\prime}, \theta^{\prime}\rangle=X^{\prime\mu}|X^{\prime}, \theta^{\prime}\rangle
\qquad \mbox{and} \qquad \hat{\theta}^{\mu\nu}|X^{\prime},
\theta^{\prime}\rangle=\theta^{\prime\mu\nu}|X^{\prime},\theta^{\prime}\rangle \; ,
\end{eqnarray}
and
\begin{eqnarray}\label{eqautovalorespmupimunu}
\hat{p}_{\mu}|p^{\prime},k^{\prime} \rangle
= p^{\prime}_{\mu}|p^{\prime},k^{\prime} \rangle
\hspace{0.5cm} , \hspace{0.5cm} \quad
\hat{k}_{\mu\nu}|p^{\prime},k^{\prime} \rangle
= k^{\prime}_{\mu\nu}|p^{\prime},k^{\prime} \rangle \; .
\end{eqnarray}

The wave function $\phi(X^{\prime}, \theta^{\prime})=\langle X^{\prime}, \theta^{\prime}|\phi\rangle$
represents the physical state $|\phi\rangle$ in the coordinate basis defined above.
This wave function satisfies some wave equation that can be derived from an action,
through a variational principle, as usual. In \cite{Amorim4}, the author constructed directly
an ordinary relativistic free quantum theory.
It was assumed that the physical states are annihilated
by the mass-shell condition
\begin{eqnarray}\label{reldisp}
\left(\hat{p}_{\mu}\hat{p}^{\mu}-m^2\right)|\phi \rangle = 0 \; ,
\end{eqnarray}
demonstrated through the Casimir operator $C_{1}=\hat{p}_{\mu}\hat{p}^{\mu}$ (for more algebraic
details see \cite{Amorim4}). It is easy to see that in the coordinate representation, this originates
the NC Klein-Gordon equation. Condition (\ref{reldisp}) selects the physical states that must be invariant under
gauge transformations. To treat the NC case, let us assume that the second mass-shell condition
\begin{eqnarray}\label{reldisppi}
\left( \hat{k}_{\mu\nu}\hat{k}^{\mu\nu}-\Delta^2 \right)|\phi\rangle = 0 \; ,
\end{eqnarray}
and must be imposed on the physical states, where $\Delta$ is some constant with dimension $M^4$,
which sign and value can be defined if $k$ is spacelike, timelike or null.
Analogously the Casimir invariant is $C_{2}=\hat{k}_{\mu\nu}\hat{k}^{\mu\nu}$, demonstrated
the validity of (\ref{reldisp}) (see \cite{Amorim4} for details).

Both equations (\ref{reldisp}) and ({\ref{reldisppi}) permit us to construct a general
expression for the plane wave solution such as \cite{Amorim4}
\begin{eqnarray}\label{phiXtheta}
\phi(x^{\prime},\theta^{\prime}):= \langle X^{\prime},\theta^{\prime}|\phi\rangle
=\int \frac{d^{4}p}{(2\pi)^{4}} \frac{d^{6}k}{(2\pi\lambda^{-2})^{6}} \, \widetilde{\phi}(p,k^{\mu\nu})
\, \exp \left( ip_{\mu}x^{\prime\mu}+\frac i2 k_{\mu\nu} \theta^{\prime \mu\nu} \right) \; ,
\end{eqnarray}
where $p^2\,-\,m^2=0$ and $k^2\,-\,\Delta^2=0$, and we have used that $p\cdot X=p \cdot x$. The length
$\lambda^{-2}$ is introduced conveniently in the $k$-integration to maintain the field with dimension
of length inverse. Consequently, the $k$-integration stays dimensionless.

In coordinate representation, the operators $(\hat{p},\hat{k})$ are written in terms of the
derivatives
\begin{eqnarray}\label{repcoordinateppi}
\hat{p}_{\mu} \rightarrow -i\p_{\mu}
\hspace{0.6cm} \mbox{and} \hspace{0.6cm}
\hat{k}_{\mu\nu} \rightarrow -i\frac{\p}{\p \theta^{\mu\nu}} \; ,
\end{eqnarray}
and consequently, both (\ref{reldisp}) and (\ref{reldisppi}) are combined into a single equation to give
the Klein-Gordon equation in the DFR space for the scalar field $\phi$
\begin{eqnarray}\label{NCKG}
\left(\Box +\lambda^2\Box_\theta+m^2\right)\phi(x,\theta)=0 \; ,
\end{eqnarray}
where we have defined $\Box_{\theta}=\frac 12\,\p^{\mu\nu}\,\p_{\mu\nu}$
and $\p_{\mu\nu}=\frac{\p}{\p \theta^{\mu\nu}}$, with $\eta^{\mu\nu}=\mbox{diag}(1,-1,-1,-1)$.
Substituting the wave plane solution (\ref{phiXtheta}), we obtain the mass invariant
\begin{eqnarray}\label{MassInv}
p^{2}+\frac{\lambda^{2}}{2}k_{\mu\nu}k^{\mu\nu}=m^2 \; ,
\end{eqnarray}
where $\lambda$ is a parameter with dimension of length defined before,
as the Planck length. We define the components of the $k$-momentum
$k^{\mu\nu}=(-{\bf k},-\widetilde{{\bf k}})$ and $k_{\mu\nu}=({\bf k},\widetilde{{\bf k}})$,
to obtain the DFR$^*$ dispersion relation
\begin{eqnarray}\label{RelDispDFRA}
\omega({\bf p},{\bf k},\widetilde{{\bf k}})=\sqrt{{\bf p}^{2}
+\lambda^{2}\left({\bf k}^{2}+\widetilde{{\bf k}}^{2}\right)+m^2} \; ,
\end{eqnarray}
in which $\widetilde{k}_{i}$ is the dual vector of the components $k_{ij}$,
that is, $k_{ij}=\epsilon_{ijk}\widetilde{k}_{k}$ $(i,j,k=1,2,3,)$.
It is easy to see that, promoting the limit $\lambda \rightarrow 0$ in Eq. (\ref{RelDispDFRA})
causes the recovering of commutativity \cite{EMCAbreuMJNeves2012}.

To propose the action of a scalar field we need to define the Weyl representation for DFR operators.
It is given by the mapping
\begin{eqnarray}\label{mapweyl}
\hat{{\cal W}}(f)(\hat{x},\hat{\theta})=\int\frac{d^{4}p}{(2\pi)^{4}}
\frac{d^{6}k}{(2\pi\lambda^{-2})^{6}} \;
\widetilde{f}(p,k^{\mu\nu}) \; e^{ip \cdot \hat{x}+\frac{i}{2} k \cdot \hat{\theta}} \; ,
\end{eqnarray}
in which $(\hat{x},\hat{\theta})$ are the position operators satisfying the DFR$^*$ algebra, $p_{\mu}$ and $k_{\mu\nu}$
are the conjugated momentum of the coordinates $x^{\mu}$ and $\theta^{\mu\nu}$ , respectively.
The Weyl symbol provides a map from the operator algebra to the algebra of functions
equipped with a star-product, via the Weyl-Moyal correspondence
\begin{eqnarray}\label{WeylMoyal}
\hat{f}(\hat{x},\hat{\theta}) \; \hat{g}(\hat{x},\hat{\theta})
\hspace{0.3cm} \leftrightarrow \hspace{0.3cm}
f(x,\theta) \star g(x,\theta) \; ,
\end{eqnarray}
and the star-product turns out to be the same as in the usual NC case
\begin{eqnarray}\label{ProductMoyal}
\left. f(x,\theta) \star g(x,\theta) =
e^{\frac{i}{2}\theta^{\mu\nu}\partial_{\mu}\partial^{\prime}_{\nu}}
f(x,\theta) g(x^{\prime},\theta) \right|_{x^{\prime}=x} \; ,
\end{eqnarray}
for any functions $f$ and $g$. The Weyl operator (\ref{mapweyl}) has the following traces properties
\begin{eqnarray}\label{traceW}
\mbox{Tr}\left[\hat{{\cal W}}(f)\right]=\int d^{4}x \; d^{6}\theta \; W(\theta) \; f(x,\theta) \; ,
\end{eqnarray}
and for a product of $n$ functions $(f_{1},...,f_{n})$
\begin{eqnarray}\label{traceWfs}
\mbox{Tr}\left[ \hat{{\cal W}}(f_{1}) ... \hat{{\cal W}}(f_{n}) \right]=
\int d^{4}x \; d^{6}\theta \; W(\theta) \; f_{1}(x,\theta) \star ... \star f_{n}(x,\theta) \; .
\end{eqnarray}

The function $W$ is a Lorentz invariant integration-$\theta$ measure.
This weight function is introduced in the context of NC field theory to control divergences of the integration
in the $\theta$-space \cite{Carlson,Morita,Conroy2003}. It will permit us to work with series expansions
in $\theta$, {\it i.e.}, with truncated power series expansion of functions of $\theta$.
For any large $\theta_{\mu\nu}$ it falls to zero quickly so that all integrals are well defined,
in that it is assumed the normalization condition
\begin{eqnarray}\label{normalizationW}
\langle {\bf 1} \rangle = \int d^{6}\theta \; W(\theta) = 1 \; .
\end{eqnarray}
The function $W$ should be an even function of
$\theta$, that is, $W(-\theta)=W(\theta)$,
and consequently it implies that
\begin{eqnarray}\label{intWtheta}
\langle \theta^{\mu\nu} \rangle = \int d^{6} \theta \; W(\theta) \; \theta^{\mu\nu} = 0 \; .
\end{eqnarray}
The non trivial integrals are expressed in terms of the invariant
\begin{eqnarray}\label{theta2n}
\langle \theta^{2 n} \rangle=\int d^{6}\theta \; W(\theta) \; (\theta_{\mu\nu}\theta^{\mu\nu})^{n}
\; , \; \mbox{with} \; n \in {\mathbb Z}_{+} \; ,
\end{eqnarray}
in which the normalization condition corresponding to the case $n=0$. For $n=1$, we have that
\begin{eqnarray}\label{intWtheta2}
\int d^{6}\theta \; W(\theta) \; \theta^{\mu\nu} \theta^{\rho\lambda}
=\frac{\langle \theta^{2} \rangle}{6}{\bf 1}^{[\mu\nu,\rho\lambda]} \; ,
\end{eqnarray}
where ${\bf 1}^{[\mu\nu,\rho\lambda]}:=(g^{\mu\rho}g^{\nu\lambda}-g^{\mu\lambda}g^{\nu\rho})/2$
is the identity antisymmetric on index $(\mu\nu)$ and $(\rho\lambda)$. For $n=2$, we can write that
\begin{equation}\label{intWtheta4}
\int d^{6}\theta \; W(\theta) \; \theta^{\mu\nu} \theta^{\rho\lambda} \theta^{\alpha\beta} \theta^{\gamma\sigma}
=\frac{\langle \theta^{4} \rangle}{48}\left({\bf 1}^{[\mu\nu,\rho\lambda]}{\bf 1}^{[\alpha\beta,\gamma\sigma]}
+{\bf 1}^{[\mu\nu,\alpha\beta]}{\bf 1}^{[\rho\lambda,\gamma\sigma]} + {\bf 1}^{[\mu\nu,\gamma\sigma]}
{\bf 1}^{[\rho\lambda,\alpha\beta]}  \right) \; .
\end{equation}
Using the previous properties an important integration is
\begin{eqnarray}\label{IntWthetaexp}
\int d^{6}\theta \; W(\theta) \; e^{\frac{i}{2} k_{\mu\nu} \theta^{\mu\nu}} =
e^{-\frac{\langle\theta^{2}\rangle}{48}k_{\mu\nu}k^{\mu\nu}} \; ,
\end{eqnarray}
that is easily demonstrated integrating the exponential series.
By the definition of the Moyal product (\ref{ProductMoyal}) it is 
trivial to obtain the property
\begin{eqnarray}\label{Idmoyalproduct2}
\int d^{4}x\,d^{6}\theta \; W(\theta) \; f(x,\theta) \star g(x,\theta)=
\int d^{4}x\,d^{6}\theta \; W(\theta) \; f(x,\theta) \; g(x,\theta) \; .
\end{eqnarray}
%
The physical interpretation of average of the components of $\theta^{\mu\nu}$, {\it i.e.}
$\langle \theta^{2} \rangle$, is the definition of the NC energy scale \cite{Carlson}
\begin{eqnarray}\label{LambdaNC}
\Lambda_{NC}=\left(\frac{12}{\langle \theta^{2} \rangle} \right)^{1/4}=\frac{1}{\lambda} \; ,
\end{eqnarray}
in which $\lambda$ is the fundamental length scale
that emerging in the KLEIN-GORDON equation (\ref{NCKG}), and in the dispersion relation
(\ref{RelDispDFRA}). This approach has the advantage that it is not need to
specify the form of the function $W$, at least for lowest-order processes.
The study of Lorentz-invariant noncommutative QED,
as Bhabha scattering, dilepton and diphoton production
to LEP data led the authors of \cite{Conroy2003,Carone} to the bound
\begin{eqnarray}\label{boundLambda}
\Lambda_{NC} > 160 \; GeV \; \; 95 \% \; C.L. \; .
\end{eqnarray}

After the discussion of the $W$-function we postulate the completeness relations
\begin{eqnarray}\label{completezarelXtheta}
\int d^{4}x^{\prime} d^{6}\theta^{\prime} \, W(\theta^{\prime}) |X^{\prime},\theta^{\prime} \rangle
\langle X^{\prime},\theta^{\prime} | ={\bf 1} \; ,
\end{eqnarray}
and
\begin{eqnarray}\label{completezarelpkmunu}
\int \frac{d^{4}p^{\prime}}{(2\pi)^{4}} \; \frac{d^{6}k^{\prime}}{(2\pi)^{6}} \; |p^{\prime},k^{\prime} \rangle
\langle p^{\prime},k^{\prime} | ={\bf 1} \; .
\end{eqnarray}
Using the previous completeness relations and the integral (\ref{IntWthetaexp}), we can obtain
\begin{eqnarray}\label{ortognalityrelationXtheta}
\langle X^{\prime},\theta^{\prime} |X^{\prime\prime},\theta^{\prime\prime} \rangle =
\delta^{(4)}\!\left(x^{\prime}-x^{\prime\prime}\right)W^{-1}(\theta^{\prime})\delta^{(6)}\!
\left(\theta^{\prime}-{\bf \theta}^{\prime\prime}\right) \; ,
\end{eqnarray}
and
\begin{eqnarray}\label{ortognalityrelationppi}
\langle p, k |p^{\prime}, k^{\prime} \rangle =
(2\pi)^{4}\delta^{(4)}\!\left(p-p^{\prime}\right)
e^{-\frac{\lambda^{4}}{4}\left(k_{\mu\nu}-k_{\mu\nu}^{\prime}\right)^{2}} \; ,
\end{eqnarray}
where we have used $\langle \theta^{2} \rangle=12\lambda^{4}$ from (\ref{LambdaNC}),
and the matrix elements
\begin{eqnarray}\label{XthetapmuXtheta}
\langle X^{\prime},\theta^{\prime}|\hat{p}_{\mu}| X^{\prime\prime},\theta^{\prime\prime} \rangle \,=
\,-\,i\,\partial^{\prime}_{\mu}\,\delta^{(4)}(x^{\prime}-x^{\prime\prime})
W^{-1}(\theta^{\prime})\delta^{(6)}(\theta^{\prime}-\theta^{\prime\prime})
\; ,
\end{eqnarray}
and
\begin{eqnarray}\label{XthetapimunuXtheta}
\langle X^{\prime},\theta^{\prime}|\hat{k}_{\mu\nu}| X^{\prime\prime},\theta^{\prime\prime}\rangle \,
=\delta^{(4)}(x^{\prime}-x^{\prime\prime})(\,-\,i\,)
\frac{\p}{\p \theta^{\prime \mu\nu}}\left( W^{-1}(\theta^{\prime})
\delta^{(6)}(\theta^{\prime}-\theta^{\prime\prime})  \right) \; ,
\end{eqnarray}
that confirms the differential representation of (\ref{repcoordinateppi}).
The result (\ref{ortognalityrelationppi}) reveals that the
canonical momentum $k_{\mu\nu}$ associated to $\theta^{\mu\nu}$ is not conserved
due to introduction of the $W$-function.
Mathematically speaking, in (\ref{ortognalityrelationppi}) we can see that at first sight there is a momentum conservation problem at the vertice.  It will be clarified in the future.

\section{The action of $\phi^{4}\star$ model and Feynman rules}
\renewcommand{\theequation}{3.\arabic{equation}}
\setcounter{equation}{0}

The DFR$^*$ action of a scalar field with quartic self-interaction $\phi^{4}\star$, defined at the NC spacetime is given by
\begin{equation}\label{actionscalarstar}
S(\phi)=\int d^{4}x \,d^{6}\theta \, W(\theta) \left( \frac{1}{2} \partial_{\mu}\phi \star \partial^{\mu}\phi
+\frac{\lambda^{4}}{4} \partial_{\mu\nu}\phi \star \partial^{\mu\nu}\phi -\frac{1}{2}m^2\phi \star \phi -\frac{g}{4!} \phi \star \phi \star \phi \star \phi \right) \; ,
\end{equation}
where $g$ is a constant coupling, and using the identity (\ref{Idmoyalproduct2}), this action is reduced to
\begin{eqnarray}\label{actionscalar}
S(\phi)=\int d^{4}x \, \frac{1}{2} \left[ \phantom{\frac{1}{2}} \!\!\!\!\!  \left(\partial_{\mu}\phi\right)^{2}
+\frac{\lambda^{4}}{2} \left(\partial_{\mu\nu}\phi\right)^{2} -m^2\phi^{2} \right]
-\frac{g}{4!} \int d^{4}x \,d^{6}\theta \, W(\theta) \left(\phi \star \phi\right)^{2} \; .
\end{eqnarray}
Using the Fourier transform, the free action in momentum space gives us the Feynman propagator \cite{EMCAbreuMJNeves2013}
\begin{eqnarray}\label{propagadorlivrek}
\Delta_{F}(p;k_{\mu\nu},k_{\mu\nu}^{\prime})=(2\pi)^{4}\delta^{4}\left(p^{\prime}+p\right)
\frac{i\, e^{-\frac{\lambda^{4}}{4}\left(k_{\mu\nu}+k_{\mu\nu}^{\prime}\right)^{2}}}
{p^{2}+\frac{\lambda^{4}}{2}k_{\mu\nu}^{\prime}k^{\mu\nu}-m^{2}+i\varepsilon}  \; .
\end{eqnarray}
The influence of the non-commutativity via Moyal's product is in the interaction term $\phi^{4}\star$ of (\ref{actionscalar}).
To obtain the convenient Feynman rule for the vertex, we write the interaction term of (\ref{actionscalar})
in the Fourier space
\begin{eqnarray}\label{actionintmomentumdthetaw}
S_{int}(\widetilde{\phi})=-\frac{g}{4!}\int \prod_{i=1}^{4} \frac{d^{4}p_{i}}{(2\pi)^{4}}
\, \frac{d^{6}k_{i}}{(2\pi)^{6}}
\, \widetilde{\phi}(p_{i},k_{i\mu\nu})
(2\pi)^{4} \delta^{(4)}\left(p_{1}+p_{2}+p_{3}+p_{4} \right)
\, \times
\nonumber \\
\times \, \int d^{6}\theta \, W(\theta)
\,F(p_{1},p_{2},p_{3},p_{4}) \,\, \mbox{e}^{\frac{i}{2}\left(k_{1\mu\nu}+k_{2\mu\nu}+k_{3\mu\nu}+k_{4\mu\nu}\right)\theta^{\mu\nu}} \; ,
\end{eqnarray}
where $\widetilde{\phi}$ is Fourier transform of $\phi$ and $F$ is a function totally symmetric that changes the momenta $(p_{1},p_{2},p_{3},p_{4})$ as
\begin{eqnarray}\label{F}
F(p_{1},p_{2},p_{3},p_{4})=\frac{1}{3}\left[ \cos \left(\frac{p_{1} \wedge p_{2} }{2}\right) \cos \left(\frac{p_{3} \wedge p_{4} }{2}\right)
+\cos \left(\frac{p_{1} \wedge p_{3} }{2}\right)\cos \left(\frac{p_{2} \wedge p_{4} }{2}\right)
\right. \nonumber \\
\left. +\cos \left(\frac{p_{1} \wedge p_{4} }{2}\right) \cos \left(\frac{p_{2} \wedge p_{3} }{2}\right) \right] \; .
\end{eqnarray}
The symbol $\wedge$ means the product $p_{i} \wedge p_{j}=\theta^{\mu\nu}p_{\mu i} p_{\nu j}$,
if $i \neq j$ $(i,j=1,2,3,4)$, and $p_{i} \wedge p_{j}=0$, if $i=j$. The $\theta$-integral of
(\ref{actionintmomentumdthetaw}) can be calculated with help of (\ref{IntWthetaexp}) and after some manipulations we obtain
\begin{eqnarray}\label{actionintmomentum}
S_{int}(\widetilde{\phi})=-\frac{g}{6 \times 4!}\int \prod_{i=1}^{4} \frac{d^{4}p_{i}}{(2\pi)^{4}}
\frac{d^{6}k_{i}}{(2\pi)^{6}}
\,\widetilde{\phi}(p_{i},k_{i\mu\nu}) \, (2\pi)^{4} \delta^{(4)}\left(p_{1}+p_{2}+p_{3}+p_{4} \right)
\, \times
\nonumber \\
\times \,
e^{-\frac{\lambda^{4}}{4}k_{s\mu\nu}k_{s}^{\mu\nu}} \, \left\{
e^{-\frac{\lambda^{4}}{16}\left(p_{1\mu}p_{2\nu}-p_{1\nu}p_{2\mu}+p_{3\mu}p_{4\nu}-p_{3\nu}p_{4\mu}\right)^{2}}
\cosh\left[\frac{\lambda^{4}}{2}k_{s}^{\mu\nu}\left(p_{1\mu}p_{2\nu}+p_{3\mu}p_{4\nu} \right)\right]
\right. \nonumber \\
\left.
+e^{-\frac{\lambda^{4}}{16}(p_{1\mu}p_{2\nu}-p_{1\nu}p_{2\mu}-p_{3\mu}p_{4\nu}+p_{3\nu}p_{4\mu})^{2}}
\cosh\left[\frac{\lambda^{4}}{2}k_{s}^{\mu\nu}\left(p_{1\mu}p_{2\nu}-p_{3\mu}p_{4\nu} \right)\right]
\right. \nonumber \\
\left.
+e^{-\frac{\lambda^{4}}{16}(p_{1\mu}p_{3\nu}-p_{1\nu}p_{3\mu}+p_{2\mu}p_{4\nu}-p_{2\nu}p_{4\mu})^{2}}
\cosh\left[\frac{\lambda^{4}}{2}k_{s}^{\mu\nu}\left(p_{1\mu}p_{3\nu}+p_{2\mu}p_{4\nu} \right)\right]
\right. \nonumber \\
\left.
+e^{-\frac{\lambda^{4}}{16}(p_{1\mu}p_{3\nu}-p_{1\nu}p_{3\mu}-p_{2\mu}p_{4\nu}+p_{2\nu}p_{4\mu})^{2}}
\cosh\left[\frac{\lambda^{4}}{2}k_{s}^{\mu\nu}\left(p_{1\mu}p_{3\nu}-p_{2\mu}p_{4\nu} \right)\right]
\right. \nonumber \\
\left.
+e^{-\frac{\lambda^{4}}{16}(p_{1\mu}p_{4\nu}-p_{1\nu}p_{4\mu}+p_{2\mu}p_{3\nu}-p_{2\nu}p_{3\mu})^{2}}
\cosh\left[\frac{\lambda^{4}}{2}k_{s}^{\mu\nu}\left(p_{1\mu}p_{4\nu}+p_{2\mu}p_{3\nu} \right)\right]
\right. \nonumber \\
\left.
+e^{-\frac{\lambda^{4}}{16}(p_{1\mu}p_{4\nu}-p_{1\nu}p_{4\mu}-p_{2\mu}p_{3\nu}+p_{2\nu}p_{3\mu})^{2}}
\cosh\left[\frac{\lambda^{4}}{2}k_{s}^{\mu\nu}\left(p_{1\mu}p_{4\nu}-p_{2\mu}p_{3\nu} \right)\right]
\right\} \, . \hspace{0.5cm}
\end{eqnarray}
The vertex $V^{(4)}$ of the $\phi^{4}\star$ interaction in the momentum space is given by
\begin{eqnarray}\label{VertexV4k}
V^{(4)}(p_{1},\cdot\cdot\cdot,p_{4};k_{1\mu\nu},\cdot\cdot\cdot,k_{4\mu\nu})=-\frac{g}{6}
(2\pi)^{4} \delta^{(4)}\left(p_{1}+p_{2}+p_{3}+p_{4} \right) \, \times
\hspace{1.2cm}\nonumber \\
\times e^{-\frac{\lambda^{4}}{4}k_{s\mu\nu}k_{s}^{\mu\nu}} \,   \left\{
e^{-\frac{\lambda^{4}}{16}\left(p_{1\mu}p_{2\nu}-p_{1\nu}p_{2\mu}+p_{3\mu}p_{4\nu}-p_{3\nu}p_{4\mu}\right)^{2}}
\cosh\left[\frac{\lambda^{4}}{2}k_{s}^{\mu\nu}\left(p_{1\mu}p_{2\nu}+p_{3\mu}p_{4\nu} \right)\right]
\right. \nonumber \\
\left.
+e^{-\frac{\lambda^{4}}{16}(p_{1\mu}p_{2\nu}-p_{1\nu}p_{2\mu}-p_{3\mu}p_{4\nu}+p_{3\nu}p_{4\mu})^{2}}
\cosh\left[\frac{\lambda^{4}}{2}k_{s}^{\mu\nu}\left(p_{1\mu}p_{2\nu}-p_{3\mu}p_{4\nu} \right)\right]
\right. \nonumber \\
\left.
+e^{-\frac{\lambda^{4}}{16}(p_{1\mu}p_{3\nu}-p_{1\nu}p_{3\mu}+p_{2\mu}p_{4\nu}-p_{2\nu}p_{4\mu})^{2}}
\cosh\left[\frac{\lambda^{4}}{2}k_{s}^{\mu\nu}\left(p_{1\mu}p_{3\nu}+p_{2\mu}p_{4\nu} \right)\right]
\right. \nonumber \\
\left.
+e^{-\frac{\lambda^{4}}{16}(p_{1\mu}p_{3\nu}-p_{1\nu}p_{3\mu}-p_{2\mu}p_{4\nu}+p_{2\nu}p_{4\mu})^{2}}
\cosh\left[\frac{\lambda^{4}}{2}k_{s}^{\mu\nu}\left(p_{1\mu}p_{3\nu}-p_{2\mu}p_{4\nu} \right)\right]
\right. \nonumber \\
\left.
+e^{-\frac{\lambda^{4}}{16}(p_{1\mu}p_{4\nu}-p_{1\nu}p_{4\mu}+p_{2\mu}p_{3\nu}-p_{2\nu}p_{3\mu})^{2}}
\cosh\left[\frac{\lambda^{4}}{2}k_{s}^{\mu\nu}\left(p_{1\mu}p_{4\nu}+p_{2\mu}p_{3\nu} \right)\right]
\right. \nonumber \\
\left.
+e^{-\frac{\lambda^{4}}{16}(p_{1\mu}p_{4\nu}-p_{1\nu}p_{4\mu}-p_{2\mu}p_{3\nu}+p_{2\nu}p_{3\mu})^{2}}
\cosh\left[\frac{\lambda^{4}}{2}k_{s}^{\mu\nu}\left(p_{1\mu}p_{4\nu}-p_{2\mu}p_{3\nu} \right)\right]
\right\}
\; ,
\end{eqnarray}
where we have defined the sum of the $k$-momenta as
\begin{eqnarray}\label{ks}
k_{s}^{\mu\nu}=k_{1}^{\mu\nu}+k_{2}^{\mu\nu}+k_{3}^{\mu\nu}+k_{4}^{\mu\nu} \; .
\end{eqnarray}
The expressions (\ref{propagadorlivrek}) and (\ref{VertexV4}) are the Feynman rules of the
$\phi^{4}\star$ DFRA model. For radiative corrections of the perturbative series, 
lines and vertex are represented in the momentum space by those expressions. Clearly,
the vertex expression shows that the external total momentum associated to extra dimension
$\theta$ is not conserved, that is, $k_{s\mu\nu} \neq 0$, while the total usual momentum $p^{\mu}$
is conserved.   But notice that we are stating that the physical momentum is not conserved in $\theta$-space.   
We are talking about the vertex expression for the momentum.


\section{RADIATIVE CORRECTIONS}
\renewcommand{\theequation}{4.\arabic{equation}}
\setcounter{equation}{0}

In this section we will analyze carefully the radiative corrections that are affected by the NCY property of $\theta$-space.  However, we have to remember that the $k_{\mu\nu}=0$ case concerns the DFR spacetime.  The objective here is to compare with DFR$^*$ phase-space where $k_{\mu\nu} \neq 0$.  This will be carried out in the next section.

\subsection{The self-energy propagator when $k_{\mu\nu}=0$}

Now we use the Feynman rules in momentum space (\ref{propagadorlivrek}) and (\ref{VertexV4k})
to obtain the radiative corrections to the propagator and vertex of the $\phi^{4}\star$
model. For simplify we begin analyzing the divergences at the one loop approximation when the $\theta$ conjugate momentum 
$k_{\mu\nu}=0$ in the Feynman rules of the previous section. The propagator is analogous to the usual
NC commutative case
\begin{eqnarray}\label{propagadorlivre}
\Delta_{F}(p;k_{\mu\nu}=0)=\frac{i}{p^{2}-m^{2}+i\varepsilon} \; ,
\end{eqnarray}
and the vertex expression is reduced to
\begin{eqnarray}\label{VertexV4}
V^{(4)}(p_{1},\cdot\cdot\cdot,p_{4};k_{\mu\nu}=0)=-\frac{g}{6}
(2\pi)^{4} \delta^{(4)}\left(p_{1}+p_{2}+p_{3}+p_{4} \right) \times
\hspace{1.2cm}\nonumber \\
\times \left[
e^{-\frac{\lambda^{4}}{16}\left(p_{1\mu}p_{2\nu}-p_{1\nu}p_{2\mu}+p_{3\mu}p_{4\nu}-p_{3\nu}p_{4\mu}\right)^{2}}
+e^{-\frac{\lambda^{4}}{16}(p_{1\mu}p_{2\nu}-p_{1\nu}p_{2\mu}-p_{3\mu}p_{4\nu}+p_{3\nu}p_{4\mu})^{2}}
\right. \nonumber \\
\left.
+e^{-\frac{\lambda^{4}}{16}(p_{1\mu}p_{3\nu}-p_{1\nu}p_{3\mu}+p_{2\mu}p_{4\nu}-p_{2\nu}p_{4\mu})^{2}}
+e^{-\frac{\lambda^{4}}{16}(p_{1\mu}p_{3\nu}-p_{1\nu}p_{3\mu}-p_{2\mu}p_{4\nu}+p_{2\nu}p_{4\mu})^{2}}
\right. \nonumber \\
\left.
+e^{-\frac{\lambda^{4}}{16}(p_{1\mu}p_{4\nu}-p_{1\nu}p_{4\mu}+p_{2\mu}p_{3\nu}-p_{2\nu}p_{3\mu})^{2}}
+e^{-\frac{\lambda^{4}}{16}(p_{1\mu}p_{4\nu}-p_{1\nu}p_{4\mu}-p_{2\mu}p_{3\nu}+p_{2\nu}p_{3\mu})^{2}}
\right] \; .
\end{eqnarray}
We just restrict the calculation to the one loop approximation. The first non-trivial correction
to the scalar field propagator comes from the diagram

\begin{figure}[!h]
\begin{center}
\newpsobject{showgrid}{psgrid}{subgriddiv=1,griddots=10,gridlabels=6pt}
\begin{pspicture}(-0.5,3)(9,4.8)
\psset{arrowsize=0.18 2}
\psset{unit=1}
%
%
%
%
\psarc[linewidth=.2mm](3,4){1}{0}{360}
\psline[linewidth=.2mm]{->}(1,3)(1.9,3)
%
\psarc[linewidth=.2mm]{<-}(3,4){1}{0}{1}
\psarc[linewidth=.2mm]{->}(3,4){1}{0}{190}
\psline[linewidth=.2mm](1.7,3)(3,3)
\psline[linewidth=.2mm]{-}(3,3)(4.2,3)
\psline[linewidth=.2mm]{<-}(4,3)(5,3)
%
%
\rput(1.5,2.5){\large$p$}
\rput(4.3,2.5){\large$-p$}
\rput(1.6,4){\large$q$}
\rput(4.3,4){\large$q$}
%
%
\psarc[linewidth=.2mm](8,4){1}{0}{360}
\psline[linewidth=.2mm]{->}(6,3)(7,3)
%
\psarc[linewidth=.2mm]{<-}(8,4){1}{0}{1}
\psarc[linewidth=.2mm]{->}(8,4){1}{0}{190}
\psline[linewidth=.2mm](6.8,3)(8,3)
\psline[linewidth=.2mm]{-}(8,3)(9.2,3)
\psline[linewidth=.2mm]{<-}(9,3)(10,3)
%
%
\psline[linewidth=.5mm,linestyle=dashed](7.5,3.5)(8.5,2.5)
\psline[linewidth=.5mm,linestyle=dashed](8.5,3.5)(7.5,2.5)
%
%
%
\rput(6.5,2.5){\large$p$}
\rput(9.3,2.5){\large$-p$}
\rput(6.6,4){\large$q$}
\rput(9.3,4){\large$q$}
%
\rput(-1,3.1){\Large$\Sigma_{1}(m^{2},p)=$}
\rput(5.5,3.1){\large$+$}
\end{pspicture}
\end{center}
\end{figure}

\noindent
where
\begin{eqnarray}\label{Sigma1}
\Sigma_{1}(m^{2},p)=\frac{1}{2}\int \frac{d^{4}p_{1}}{(2\pi)^{4}}
\frac{d^{4}p_{2}}{(2\pi)^{4}}\frac{d^{4}p_{3}}{(2\pi)^{4}}\frac{d^{4}p_{4}}{(2\pi)^{4}}
\frac{i}{p_{2}^{2}-m^{2}+i\varepsilon} \times
\nonumber \\
\times \, (2\pi)^{4}\delta^{(4)}(p_{2}+p_{3}) (2\pi)^{4}\delta^{(4)}(p_{1}-p)
V^{(4)}(p_{1},p_{2},p_{3},p_{4}) \; ,
\hspace{0.5cm}
\end{eqnarray}
and after some calculation we have that
\begin{eqnarray}\label{Sigma1}
\Sigma_{1}(m^{2},p)=-\frac{g}{6}\int
\frac{d^{4}q}{(2\pi)^{4}}
\frac{i}{q^{2}-m^{2}+i\varepsilon}\left[
2+e^{-\frac{\lambda^{4}}{4}(q_{\mu}p_{\nu}-p_{\mu}q_{\nu})^{2}}
\right] \; .
\hspace{0.8cm}
\end{eqnarray}
\noindent where we can see the NC factor $\lambda$ and the contribution of NCY.

The previous expression suggests us to write it as
\begin{eqnarray}\label{Sigma1pn+np}
\Sigma_{1}(m^{2},p)=\Sigma_{1(pn)}(m^{2})+\Sigma_{1(np)}(m^{2},p) \; ,
\end{eqnarray}
where $\Sigma_{1(pn)}$ is the self-energy of the planar $(pn)$ diagram
\begin{eqnarray}\label{Sigma1pn}
\Sigma_{1(pn)}(m^{2})= -\frac{g}{6}\int
\frac{d^{4}q}{(2\pi)^{4}}
\frac{i}{q^{2}-m^{2}+i\varepsilon} \; ,
\end{eqnarray}
while $\Sigma_{1(np)}$ is the non-planar $(np)$ diagram
\begin{eqnarray}\label{Sigma1np}
\Sigma_{1(np)}(m^{2},p)= -\frac{g}{3}\int
\frac{d^{4}q}{(2\pi)^{4}}
\frac{i}{q^{2}-m^{2}+i\varepsilon} \,\, e^{-\frac{\lambda^{4}}{2}\left(q^2p^2-\left(q\cdot p\right)^{2}\right)} \; .
\end{eqnarray}
\noindent where from Eqs. (\ref{Sigma1pn}) and (\ref{Sigma1np}) we can see that the non-planar (np) diagram is the one affected by NCY, i.e., we have a non-planar diagram in DFR$^*$ phase-space.

The expression planar is very similar to the self-energy of the $\phi^{4}$ commutative at the one loop
approximation. It has an integral that diverges in the ultraviolet limit, that is, when
$|{\bf k}| \rightarrow \infty$. For future convenience, we use the method of Schwinger's parametrization
to evaluate the integrals (\ref{Sigma1pn}) and (\ref{Sigma1np}), in which it is introduced the integral
\begin{eqnarray}\label{SchwingerIntds}
\frac{i}{q^{2}-m^{2}+i\varepsilon}
=\int_{0}^{\infty} ds \,\,\, e^{is\;\left(q^{2}-m^{2}+i\varepsilon\right)} \; ,
\end{eqnarray}
where $s$ is the proper-time parameter that has length to the squared dimension.
The self-energy planar term is easily calculated to give the $s$-integration
\begin{eqnarray}\label{Sigma1pnIntds}
\Sigma_{1(pn)}(m^{2})= \frac{ig}{96\pi^{2}}
\int_{0}^{\infty} \frac{ds}{s^{2}} \,\,\, e^{-is\;\left(m^{2}-i\varepsilon\right)}
  \; .
\end{eqnarray}
This $s$-integral evidently diverges when $s=0$.   Then we introduce a cutoff function regulator
given by $e^{-1/i4s\Lambda^{2}}$, where $\Lambda$ is a momenta cutoff parameter
\begin{eqnarray}\label{Sigma1pnIntdsCutLambda}
\Sigma_{1(pn)}(m^{2},\Lambda)= \frac{ig}{96\pi^{2}}
\int_{0}^{\infty} \frac{ds}{s^{2}} \,\,\, e^{-is\;\left(m^{2}-i\varepsilon\right)-\left(i4s\Lambda^{2}\right)^{-1}} \; ,
\end{eqnarray}
in which the divergence is naturally recovered in the limit
$\Lambda \rightarrow \infty$, and the function regulator is one.
Using the Handbook integral \cite{Gradshteyn00}, we obtain the
regularized function
\begin{eqnarray}\label{Sigma1pnCutLambdaBessel}
\Sigma_{1(pn)}(m^{2},\Lambda/m)= -\frac{gm^{2}}{24\pi^{2}} \, \frac{\Lambda}{m} \, K_{1}\left(\frac{m}{\Lambda}\right)\; ,
\end{eqnarray}
where $K_{1}$ is the Bessel function of the second kind. When $\Lambda/m \gg 1$,
we expand the Bessel function to obtain the result
\begin{eqnarray}\label{Sigma1pnCutLambdaBesselExpand}
\Sigma_{1(pn)}(m^{2},\Lambda/m) \approx -\frac{gm^{2}}{24\pi^{2}} \, \frac{\Lambda^{2}}{m^{2}}-\frac{gm^{2}}{48\pi^{2}} \ln\left(\frac{m}{\Lambda} \right) \; ,
\end{eqnarray}
that has a quadratic divergence summed to the logarithmic one in $\Lambda \rightarrow \infty$.
In the non-planar expression (\ref{Sigma1np}), we have an exponential function in the numerator,
that suggest one to introduce the Schwinger's parametrization (\ref{SchwingerIntds}) in (\ref{Sigma1np}),
so we obtain
\begin{eqnarray}\label{Sigma1npIntds}
\Sigma_{1(np)}(m^{2},p)= -\frac{g}{3}
\int_{0}^{\infty} ds \,\, e^{-is\;\left(m^{2}-i\varepsilon\right)}
\int\frac{d^{4}q}{(2\pi)^{4}} \,\, e^{isq^{2}-\frac{\lambda^{4}}{2}\left(q^2p^2-\left(q\cdot p\right)^{2}\right)} \; .
\end{eqnarray}
These quadri-integral can be written explicitly in terms of the integrals in components $(q_{0},{\bf q})$
\begin{eqnarray}\label{Sigma1npIntdsIntdq0d3q}
\Sigma_{1(np)}(m^{2},p)= -\frac{g}{3}
\int_{0}^{\infty} ds \,\, e^{-is\;\left(m^{2}-i\varepsilon\right)}
\int \frac{d^{3}{\bf q}}{(2\pi)^{3}} \,\, e^{-\left(is-\frac{\lambda^{4}}{2}p_{\mu}p^{\mu}\right){\bf q}^{2}
+\frac{\lambda^{4}}{2}\left({\bf q}\cdot{\bf p}\right)^{2}} \,\, \times
\nonumber \\
\times \,\,\int_{-\infty}^{\infty} \frac{dq_{0}}{2\pi} \,\, e^{\left(is+\frac{\lambda^{4}}{2}{\bf p}^{2}\right)q_{0}^{2}
-\lambda^{4}p_{0}\left({\bf q} \cdot {\bf p}\right)q_{0}} \; ,
\end{eqnarray}
where the $q_{0}$-integral is a Gaussian one that can be computed to give us
\begin{eqnarray}\label{Sigma1npIntdsIntd3q}
\Sigma_{1(np)}(m^{2},p^{\mu})= -\frac{ig}{6\pi}
\int_{0}^{\infty} ds \, \left(\frac{\pi}{is+\frac{\lambda^{4}}{2}{\bf p}^{2}}\right)^{1/2} e^{-is\;\left(m^{2}-i\varepsilon\right)}
\, \times
\nonumber \\
\times \, \int \frac{d^{3}{\bf q}}{(2\pi)^{3}} \,\, e^{-\left(is-\frac{\lambda^{4}}{2}p_{\mu}p^{\mu}\right){\bf q}^{2}
+\frac{\lambda^{4}}{2}\left({\bf q}\cdot{\bf p}\right)^{2}
\left(\frac{is-\frac{\lambda^{4}}{2}p_{\mu}p^{\mu}}{is+\frac{\lambda^{4}}{2}{\bf p}^{2}} \right)} \; .
\end{eqnarray}
Using the spherical coordinates, we calculate the angular integrals to obtain the radial integral
\begin{eqnarray}\label{Sigma1pLambdaIntk}
\Sigma_{1(np)}(m^{2},p^{\mu})= -\frac{g}{6} \frac{i}{\lambda^{2}|{\bf p}|} \int_{0}^{\infty} ds \,\, \frac{e^{-is(m^{2}-i\varepsilon)}}{\left(is-\frac{\lambda^{4}}{2}p_{\mu}p^{\mu}\right)^{1/2}} \times
\hspace{0.8cm} \nonumber \\
\times \int_{0}^{\infty} \frac{dq^{2}}{(2\pi)^{2}} \,\,
e^{-\left(is-\frac{\lambda^{4}}{2}p_{\mu}p^{\mu}\right)q^{2}} \, \mbox{erfi}\left[q\,\frac{\lambda^{2}}{2}|{\bf p}|\left(\frac{is-\frac{\lambda^{4}}{2}p_{\mu}p^{\mu}}{is+\frac{\lambda^{4}}{2}{\bf p}^{2}}\right)^{\!\!1/2} \right]
\; ,
\end{eqnarray}
where $\mbox{erfi}$ is the imaginary error function \cite{Gradshteyn00}. It is not difficult to see that this integral does diverges in the $k$-integration ($\theta$-space) when $p_{\mu}p^{\mu}$ is null or time-like, that is, $p_{\mu}p^{\mu}\geq0$. In contrast, it is well defined when $p_{\mu}p^{\mu}$ is space-like, that is, $p_{\mu}p^{\mu}<0$. For this condition, we obtain
\begin{eqnarray}\label{Sigma1npsint}
\Sigma_{1(np)}(m^{2},p_{\mu}p^{\mu}) = \frac{ig}{48\pi^{2}}
\int_{0}^{\infty} \frac{ds}{s^{2}} \,\, e^{-is(m^{2}-i\varepsilon)}
\left(1+i\frac{\lambda^{4}}{2s}\,p_{\mu}p^{\mu}\right)^{-3/2} \; ,
\end{eqnarray}
in which $p_{\mu}p^{\mu}<0$. Making the $s$-integral, we obtain the result
\begin{eqnarray}\label{Sigma1nppartdiv+finite}
\Sigma_{1(np)}(m^{2},p_{\mu}p^{\mu}) = \frac{gm^{2}}{24\pi^{2}}
\frac{\sqrt{\pi}}{\lambda^{4}m^{2}p_{\mu}p^{\mu}} \, U\left(\frac{1}{2},0;-\frac{\lambda^{4}}{2}m^{2}p_{\mu}p^{\mu}\right) \; ,
\end{eqnarray}
where $U$ is a confluent hypergeometric function (Kummer's function). When we expand the previous result around the
$p_{\mu}p^{\mu} \rightarrow 0^{-}$, it emerges a ultraviolet divergence summed up to infrared divergence term
in $p^{\mu}p^{\mu}=0$
\begin{eqnarray}\label{Sigma1nppartdiv+finitep0}
\lim_{p \rightarrow 0^{-}}\Sigma_{1(np)}(m^{2},p_{\mu}p^{\mu}) = \frac{g}{12\pi^{2}}\frac{1}{\lambda^{4}p_{\mu}p^{\mu}}
-\frac{gm^{2}}{48\pi^{2}} \left[1+\gamma+\ln\left(-\frac{\lambda^{4}m^{2}}{8}p_{\mu}p^{\mu}\right)\right]
\; .
\end{eqnarray}
This result reveals itself as being a mix of ultraviolet/infrared $(UV/IR)$ divergences in DFR$^*$ NC models.
It is immediate to compare this conclusion with the usual NC case when $\theta^{\mu\nu}$
is just a real parameter
\begin{eqnarray}\label{Sigma1nppartdiv+finitep0}
\lim_{p \rightarrow 0} \Sigma_{1(np)}(m^{2},p_{\mu}p^{\mu}) = \frac{g}{96\pi^{2}} \frac{1}{\widetilde{p}_{\mu}\widetilde{p}^{\mu}}
+\frac{gm^{2}}{96\pi^{2}} \ln\left(\frac{m^{2}}{4\widetilde{p}_{\mu}\widetilde{p}^{\mu}}\right)
\; ,
\end{eqnarray}
where $\widetilde{p}_{\mu}:=\theta_{\mu\nu}p^{\nu}$. This non-planar expression also exhibits a mixing $(UV/IR)$
when $p_{\mu}p^{\mu}=0$. The two NC formalisms exhibit an analogous $(UV/IR)$ mixing.

\pagebreak

\subsection{The self-energy propagator when $k_{\mu\nu}\neq0$}

The case of $k^{\mu\nu}\neq0$ is more complicated since we must use the Feynman rules
in the momentum space (\ref{propagadorlivrek}) and (\ref{VertexV4k}). The correction
to one loop in the propagator is illustrated by the diagram
\begin{figure}[!h]
\begin{center}
\newpsobject{showgrid}{psgrid}{subgriddiv=1,griddots=10,gridlabels=6pt}
\begin{pspicture}(-0.5,2.5)(7.2,4.8)
\psset{arrowsize=0.18 2}
\psset{unit=1}
%
%
%
%
\psarc[linewidth=.2mm](3,4){1}{0}{360}
\psline[linewidth=.2mm]{->}(1,3)(1.9,3)
%
\psarc[linewidth=.2mm]{<-}(3,4){1}{0}{1}
\psarc[linewidth=.2mm]{->}(3,4){1}{0}{190}
\psline[linewidth=.2mm](1.7,3)(3,3)
\psline[linewidth=.2mm]{-}(3,3)(4.2,3)
\psline[linewidth=.2mm]{<-}(4,3)(5,3)
%
%
\rput(1.5,2.5){\large$p,k_{1\mu\nu}$}
\rput(4.5,2.5){\large$-p,k_{4\mu\nu}$}
\rput(1.3,4){\large$q,k_{2}$}
\rput(4.6,4){\large$q,k_{3}$}
%
%
\psarc[linewidth=.2mm](8,4){1}{0}{360}
\psline[linewidth=.2mm]{->}(6,3)(7,3)
%
\psarc[linewidth=.2mm]{<-}(8,4){1}{0}{1}
\psarc[linewidth=.2mm]{->}(8,4){1}{0}{190}
\psline[linewidth=.2mm](6.8,3)(8,3)
\psline[linewidth=.2mm]{-}(8,3)(9.2,3)
\psline[linewidth=.2mm]{<-}(9,3)(10,3)
%
%
\psline[linewidth=.5mm,linestyle=dashed](7.5,3.5)(8.5,2.5)
\psline[linewidth=.5mm,linestyle=dashed](8.5,3.5)(7.5,2.5)
%
%
%
\rput(6.5,2.5){\large$p,k_{1\mu\nu}$}
\rput(9.7,2.5){\large$-p,k_{4\mu\nu}$}
\rput(6.3,4){\large$q,k_{2}$}
\rput(9.6,4){\large$q,k_{3}$}
%
%
\rput(-1.8,3.1){\large$\Sigma_{1}(m^{2},p^{\mu};k_{1\mu\nu},k_{4\mu\nu})=$}
\rput(5.5,3.1){\large$+$}
\end{pspicture}
\end{center}
\end{figure}

\noindent
where the self-energy is given by
\begin{eqnarray}\label{Sigma1}
\Sigma_{1}(m^{2},p^{\mu};k_{1}+k_{4})=\frac{1}{2} \int \frac{d^{4}p_{1}}{(2\pi)^{4}}
\frac{d^{4}p_{2}}{(2\pi)^{4}}\frac{d^{4}p_{3}}{(2\pi)^{4}}\frac{d^{4}p_{4}}{(2\pi)^{4}}
\frac{d^{6}k_{2}}{(2\pi)^{6}}\frac{d^{6}k_{3}}{(2\pi)^{6}}
\frac{i\, e^{-\frac{\lambda^{4}}{4}\left(k_{2\mu\nu}+k_{3\mu\nu}\right)^{2}} }{p_{2}^{2}
+\frac{\lambda^{2}}{2}k_{2\mu\nu}k_{3}^{\mu\nu}-m^{2}+i\varepsilon} \times
\nonumber \\
\times \, (2\pi)^{4}\delta^{(4)}(p_{2}+p_{3}) (2\pi)^{4}\delta^{(4)}(p_{1}-p)
V^{(4)}(p_{1},\cdot\cdot\cdot,p_{4};k_{1\mu\nu},\cdot\cdot\cdot,k_{4\mu\nu}) \; .
\hspace{0.7cm}
\end{eqnarray}
By integrating the delta functions, the previous expression is reduced to
\begin{eqnarray}\label{Sigma1}
\Sigma_{1}(m^{2},p^{\mu};k_{1}+k_{4})=-\frac{g}{6}\int
\frac{d^{6}k_{2}}{(2\pi)^{6}}\frac{d^{6}k_{3}}{(2\pi)^{6}} \; \;
e^{-\frac{\lambda^{4}}{4}\left(k_{2\mu\nu}+k_{3\mu\nu}\right)^{2}} \, \times
\nonumber \\
\, \times \, e^{-\frac{\lambda^{4}}{4}\left(k_{1\mu\nu}+k_{2\mu\nu}+k_{3\mu\nu}+k_{4\mu\nu}\right)^{2}}
\, \int \frac{d^{4}q}{(2\pi)^{4}}
\frac{i}{q^{2}+\frac{\lambda^{2}}{2}k_{2\mu\nu}k_{3}^{\mu\nu}-m^{2}+i\varepsilon} \, \times \,
\nonumber \\
\, \times \, \left[
2+e^{-\frac{\lambda^{4}}{4}(q_{\mu}p_{\nu}-p_{\mu}q_{\nu})^{2}}\cosh\left(\lambda^{4}
(k_{1\mu\nu}+k_{2\mu\nu}+k_{3\mu\nu}+k_{4\mu\nu})q^{\mu}p^{\nu} \right)
\right] \; ,
\end{eqnarray}
where we rewrite it as a sum of a planar and non-planar parts, as we did before, so
\begin{eqnarray}\label{Sigma1pn}
\Sigma_{1(pn)}(m^{2};k_{1}+k_{4})= -\frac{g}{3}\int
\frac{d^{6}k_{2}}{(2\pi)^{6}}\frac{d^{6}k_{3}}{(2\pi)^{6}} \; \;
e^{-\frac{\lambda^{4}}{4}\left(k_{2\mu\nu}+k_{3\mu\nu}\right)^{2}} \, \times
\nonumber \\
\, \times \, e^{-\frac{\lambda^{4}}{4}\left(k_{1\mu\nu}+k_{2\mu\nu}+k_{3\mu\nu}+k_{4\mu\nu}\right)^{2}}
\, \int \frac{d^{4}q}{(2\pi)^{4}}
\frac{i}{q^{2}+\frac{\lambda^{2}}{2}k_{2\mu\nu}k_{3}^{\mu\nu}-m^{2}+i\varepsilon} \, \; ,
\end{eqnarray}
and
\begin{eqnarray}\label{Sigma1npk}
\Sigma_{1(np)}(m^{2},p^{\mu};k_{1}+k_{4})= -\frac{g}{6}\int
\frac{d^{6}k_{2}}{(2\pi)^{6}}\frac{d^{6}k_{3}}{(2\pi)^{6}} \; \;
e^{-\frac{\lambda^{4}}{4}\left(k_{2\mu\nu}+k_{3\mu\nu}\right)^{2}} \, \times
\nonumber \\
\, \times \, e^{-\frac{\lambda^{4}}{4}\left(k_{1\mu\nu}+k_{2\mu\nu}+k_{3\mu\nu}+k_{4\mu\nu}\right)^{2}}
\, \int \frac{d^{4}q}{(2\pi)^{4}}
\frac{i}{q^{2}+\frac{\lambda^{2}}{2}k_{2\mu\nu}k_{3}^{\mu\nu}-m^{2}+i\varepsilon} \, \times \,
\nonumber \\
\, \times \, e^{-\frac{\lambda^{4}}{2}\left(q^{2}p^{2}-(p \cdot q)^{2}\right)}\cosh\left[\lambda^{4}
(k_{1\mu\nu}+k_{2\mu\nu}+k_{3\mu\nu}+k_{4\mu\nu})q^{\mu}p^{\nu} \right] \; ,
\end{eqnarray}
respectively.   Notice that in this case, both planar and non-planar parts are affected by NCY.

The planar self-energy has a ultraviolet divergence in the $q$-integration, and consequently,
we use the same technique of the previous section to calculate it, so we obtain
\begin{eqnarray}\label{Sigma1pnIntdsdk}
\Sigma_{1(pn)}(m^{2};k_{1}+k_{4})= \frac{ig}{48\pi^{2}}
\int_{0}^{\infty} \frac{ds}{s^{2}} \,\, e^{-is(m^{2}-i\varepsilon)} \, \times
\hspace{1.9cm}\nonumber \\
\times \,\, \int \frac{d^{6}k_{2}}{(2\pi\lambda^{-2})^{6}}\frac{d^{6}k_{3}}{(2\pi\lambda^{-2})^{6}} \; \;
e^{-\frac{\lambda^{4}}{4}\left(k_{2\mu\nu}+k_{3\mu\nu}\right)^{2}
-\frac{\lambda^{4}}{4}\left(k_{1\mu\nu}+k_{2\mu\nu}+k_{3\mu\nu}+k_{4\mu\nu}\right)^{2}+is\frac{\lambda^{2}}{2}k_{2\mu\nu}k_{3}^{\mu\nu}}
\, \; .
\end{eqnarray}
The Gaussian integrals can be calculated to construct the planar sector as
\begin{equation}\label{Sigma1pnIntds}
\Sigma_{1(pn)}(m^{2};k_{1}+k_{4})= \frac{ig}{48\pi^{8}}\,
\int_{0}^{\infty} \frac{ds}{s^{2}} \left(\frac{\lambda^{2}}{s}\right)^{6}
\frac{e^{-is(m^{2}-i\varepsilon)}}{\left(1+i4\lambda^{2}/s\right)^{3}}
\,\, e^{-\frac{\lambda^{4}}{4}(k_{1\mu\nu}+k_{4\mu\nu})^{2}\left(\frac{1+i2\lambda^{2}/s}{1+i4\lambda^{2}/s}\right)}\,\,,
\; .
\end{equation}
and we can see that the previous result has mass dimension to the squared, which agrees with the dimension of the self energy.
This $s$-integral has a strong ultraviolet divergence when $s=0$, due to the term $s^{-8}$.  Hence, it needs a cutoff
regulator ultraviolet to turn it well defined.
Concerning the non-planar expression (\ref{Sigma1npk}), we can use the Schwinger's parametrization
to write this integral in terms of exponentials
\begin{eqnarray}\label{Sigma1npIntdsdqdk}
\Sigma_{1(np)}(m^{2},p^{\mu};k_{1}+k_{4})= -\frac{g}{6} \int_{0}^{\infty} ds \,\, e^{-is\left(m^{2}-i\varepsilon\right)}
\int \frac{d^{6}k_{2}}{(2\pi\lambda^{-2})^{6}}\frac{d^{6}k_{3}}{(2\pi\lambda^{-2})^{6}} \,\, \times
\nonumber \\
\times \,\, e^{-\frac{\lambda^{4}}{4}\left(k_{2\mu\nu}+k_{3\mu\nu}\right)^{2}
-\frac{\lambda^{4}}{4}\left(k_{1\mu\nu}+k_{2\mu\nu}+k_{3\mu\nu}+k_{4\mu\nu}\right)^{2}+is\frac{\lambda^{2}}{2}k_{2\mu\nu}k_{3}^{\mu\nu}} \, \times
\hspace{1cm} \nonumber \\
\, \times \,
\, \int \frac{d^{4}q}{(2\pi)^{4}} \,\, e^{isq^{2}-\frac{\lambda^{4}}{2}\left(q^{2}p^{2}-(p \cdot q)^{2}\right)}
\cosh\left[\lambda^{4}
\left(k_{1\mu\nu}+k_{2\mu\nu}+k_{3\mu\nu}+k_{4\mu\nu}\right)q^{\mu}p^{\nu} \right] \; .
\end{eqnarray}
where we have written the hyperbolic cosine in this expression in terms of exponentials to simplify this integral as
the sum
\begin{eqnarray}\label{CoshSumExp}
\cosh\left[\lambda^{4}
\left(k_{1\mu\nu}+k_{2\mu\nu}+k_{3\mu\nu}+k_{4\mu\nu}\right)q^{\mu}p^{\nu} \right] =
\frac{1}{2}\sum_{\sigma=-1}^{1}
e^{\sigma\,\lambda^{4}
\left(k_{1\mu\nu}+k_{2\mu\nu}+k_{3\mu\nu}+k_{4\mu\nu}\right)q^{\mu}p^{\nu}} \; ,
\end{eqnarray}
with $\sigma\neq0$.  So, we have that
\begin{eqnarray}\label{Sigma1npIntdsdqdksigma}
\Sigma_{1(np)}(m^{2},p^{\mu};k_{1}+k_{4})= -\frac{g}{12} \sum_{\sigma=-1}^{1} \int_{0}^{\infty} ds
\,\, e^{-is\left(m^{2}-i\varepsilon\right)} \, \times \,
\hspace{0.7cm} \nonumber \\
\, \times \, \int \frac{d^{4}q}{(2\pi)^{4}} \,\, e^{isq^{2}-\frac{\lambda^{4}}{2}\left(q^{2}p^{2}-(p \cdot q)^{2}\right)}
\int \frac{d^{6}k_{2}}{(2\pi\lambda^{-2})^{6}}\frac{d^{6}k_{3}}{(2\pi\lambda^{-2})^{6}} \,\, e^{-\frac{\lambda^{4}}{4}\left(k_{2\mu\nu}+k_{3\mu\nu}\right)^{2}} \, \times \,
\nonumber \\
\, \times \,\, e^{-\frac{\lambda^{4}}{4}\left(k_{1\mu\nu}+k_{2\mu\nu}+k_{3\mu\nu}+k_{4\mu\nu}\right)^{2}+is\frac{\lambda^{2}}{2}k_{2\mu\nu}k_{3}^{\mu\nu}+\sigma \lambda^{4}(k_{1\mu\nu}+k_{2\mu\nu}+k_{3\mu\nu}+k_{4\mu\nu})q^{\mu}p^{\nu}} \; .
\hspace{0.3cm}
\end{eqnarray}
%
After the calculation of the Gaussian integrals in $(k_{2},k_{3})$, we obtain
\begin{eqnarray}\label{Sigma1npIntdsdqsigma}
\Sigma_{1(np)}(m^{2},p^{\mu};k_{1}+k_{4})= -\frac{g}{12}
\sum_{\sigma=-1}^{1} \int_{0}^{\infty} ds
\, \left(\frac{\lambda^{2}}{\pi s} \right)^{6} \, \frac{e^{-is\left(m^{2}-i\varepsilon\right)}}{\left(1+i4\lambda^{2}/s\right)^{3}}
\, \times
\nonumber \\
\times \, e^{-\frac{\eta}{\sigma}\frac{\lambda^{4}}{4}\left(k_{1\mu\nu}+k_{4\mu\nu}\right)^{2}}
\int \frac{d^{4}q}{(2\pi)^{4}} \,\, e^{isq^{2}-\frac{\lambda^{4}}{2}\xi\left(q^{2}p^{2}-(p \cdot q)^{2}\right)
+\lambda^{4}\eta\,\left(k_{1\mu\nu}+k_{4\mu\nu}\right) q^{\mu}p^{\nu}}
\; ,
\end{eqnarray}
where we have defined the complex constants
\begin{eqnarray}\label{xieta}
\xi(\sigma):=1-\frac{i2\lambda^{2}/s}{1+i4\lambda^{2}/s}\sigma^{2}
\hspace{0.5cm} \mbox{and} \hspace{0.5cm}
\eta(\sigma):= \sigma \, \frac{1+i2\lambda^{2}/s}{1+i4\lambda^{2}/s} \; ,
\end{eqnarray}
in which $\xi(\sigma=\pm 1)=\left(1+i2\lambda^{2}/s\right)\left(1+i4\lambda^{2}/s\right)^{-1}$, and $\eta(\sigma=\pm 1)=\pm \xi$
that agrees with the sum defined in (\ref{CoshSumExp}).
By writing it in terms of integrals $(q_0,{\bf q})$ we have that
\begin{eqnarray}\label{Sigma1npIntdsdq0d3qsigma}
\Sigma_{1(np)}(m^{2},p^{\mu};k_{1}+k_{4})= -\frac{g}{12}
\sum_{\sigma=-1}^{1} \int_{0}^{\infty} ds
\, \left(\frac{\lambda^{2}}{\pi s} \right)^{6} \, \frac{e^{-is\left(m^{2}-i\varepsilon\right)}}{\left(1+i4\lambda^{2}/s\right)^{3}}
\, \times
\nonumber \\
\times \, e^{-\frac{\eta}{\sigma}\frac{\lambda^{4}}{4}\left(k_{1\mu\nu}+k_{4\mu\nu}\right)^{2}}
\!\!\int \frac{d^{3}{\bf q}}{(2\pi)^{3}} \, e^{-\left(is-\frac{\lambda^{4}}{2}\xi p_{\mu}p^{\mu} \right){\bf q}^{2}
+\frac{\lambda^{4}}{2} \xi \left({\bf q}\cdot{\bf p}\right)^{2}
-\lambda^{4}\eta\,p_{\mu}\left(k_{1}+k_{4}\right)^{\mu j}q^{j}}
\, \times \nonumber \\
\times \, \int_{-\infty}^{\infty}
\frac{dq_{0}}{2\pi} \,\, e^{\left(is+\frac{\lambda^{4}}{2}\xi{\bf p}^{2} \right)q_{0}^{2}
-\lambda^{4}\left[\xi \left({\bf q}\cdot{\bf p}\right) p_{0}-\eta p_{\mu}\left(k_{1}+k_{4}\right)^{\mu0}\right]q_{0}}
\; ,
\end{eqnarray}
where the $q_{0}$-integral is a Gaussian one that after computation we have that
\begin{eqnarray}\label{Sigma1npIntdsd3qsigma}
\Sigma_{1(np)}(m^{2},p^{\mu};k_{1}+k_{4})= -\frac{ig}{24\pi}
\sum_{\sigma=-1}^{1} \int_{0}^{\infty} ds
\, \left(\frac{\lambda^{2}}{\pi s} \right)^{6} \left(\frac{\pi}{is+\frac{\lambda^{4}}{2}\xi{\bf p}^{2}}\right)^{1/2}
\, \times
\nonumber \\
\times \, \frac{e^{-is\left(m^{2}-i\varepsilon\right)}}{\left(1+i4\lambda^{2}/s\right)^{3}} \, e^{-\frac{\eta}{\sigma}\frac{\lambda^{4}}{4}\left(k_{1\mu\nu}+k_{4\mu\nu}\right)^{2}
-\frac{\lambda^{8}}{4}\eta^{2}\left[p_{\mu}\left(k_{1}+k_{4}\right)^{\mu0} \right]^{2}\left(is+\frac{\lambda^{4}}{2}\xi{\bf p}^{2}\right)^{-1}}
\, \times
\nonumber \\
\times \, \int \frac{d^{3}{\bf q}}{(2\pi)^{3}} \, e^{-\left(is-\frac{\lambda^{4}}{2}\xi p_{\mu}p^{\mu} \right){\bf q}^{2}
+\frac{\lambda^{4}}{2} \xi \left({\bf q}\cdot{\bf p}\right)^{2}\left(is-\frac{\lambda^{4}}{2}\xi p_{\mu}p^{\mu}\right)
\left(is+\frac{\lambda^{4}}{2}\xi {\bf p}^{2}\right)^{-1}} \, \times
\nonumber \\
\times \,\, e^{\lambda^{4}\eta\left[\frac{\lambda^{4}}{2}\xi \, p_{\mu} \left(k_{1}+k_{4}\right)^{\mu0} \left(is+\frac{\lambda^{4}}{2}\xi{\bf p}^{2}\right)^{-1}
p_{0}p^{j}-p_{\mu}\left(k_{1}+k_{4}\right)^{\mu\,j}\right]q^{j}}
\; .
\end{eqnarray}
We have used the spherical coordinates to reduce the previous integrals to the radial integral
\begin{eqnarray}\label{Sigma1npIntdsdqsigma}
\Sigma_{1(np)}(m^{2},p^{\mu};k_{1}+k_{4})= -\frac{ig}{48}
\sum_{\sigma=-1}^{1} \int_{0}^{\infty} ds
\, \left(\frac{\lambda^{2}}{\pi s} \right)^{6} \frac{1}{\left(is-\frac{\lambda^{4}}{2}\xi p_{\mu}p^{\mu}\right)^{1/2}}
\, \times
\nonumber \\
\times \,\, \frac{e^{-is\left(m^{2}-i\varepsilon\right)}}{\left(1+i4\lambda^{2}/s\right)^{3}} \, e^{-\frac{\eta}{\sigma}\frac{\lambda^{4}}{4}\left(k_{1\mu\nu}+k_{4\mu\nu}\right)^{2}
-\frac{\lambda^{8}}{4}\eta^{2}\left[ p_{\mu}\left(k_{1}+k_{4}\right)^{\mu0} \right]^{2}
\left(is+\frac{\lambda^{4}}{2}\xi {\bf p}^{2}\right)^{-1}}
\, \times
\nonumber \\
\times \,
e^{\frac{\lambda^{4}\eta^{2}}{2\xi{\bf p}^{2}}\left|\frac{\lambda^{4}}{2} \xi \, p_{\mu} \left(k_{1}+k_{4}\right)^{\mu0} \left(is+\frac{\lambda^{4}}{2}\xi{\bf p}^{2}\right)^{-1}
p_{0}p^{j}-p_{\mu}\left(k_{1}+k_{4}\right)^{\mu\,j}\right|^{2}\left(is+\frac{\lambda^{4}}{2}\xi {\bf p}^{2}\right)\left(is-\frac{\lambda^{4}}{2}\xi p_{\mu}p^{\mu}\right)^{-1}
} \, \times
\nonumber \\
\times \, \left(\frac{2}{\lambda^{4} \xi {\bf p}^2}\right)^{1/2} \! \int_{0}^{\infty} \frac{dq^{2}}{(2\pi)^{2}} \, \, e^{-\left(is-\frac{\lambda^{4}}{2}\xi p_{\mu}p^{\mu} \right)\,q^{2}}
\, \times
\nonumber \\
\times \,
\left\{\mbox{erfi}\left[q\,\lambda^{2} |{\bf p}|\left(\frac{\xi}{2}\frac{is-\frac{\lambda^{4}}{2}\xi p_{\mu}p^{\mu}}
{is+\frac{\lambda^{4}}{2}\xi {\bf p}^{2}}\right)^{\!1/2}
\right. \right. \nonumber \\
\left. \left.
+\lambda^{2}\eta\frac{\left|\lambda^{4}\xi p_{\mu}\left(k_{1}+k_{4}\right)^{\mu 0}p_{0}p^{j}
-2p_{\mu}\left(k_{1}+k_{4}\right)^{\mu j}\left(is+\frac{\lambda^{4}}{2}\xi{\bf p}^{2}\right)\right|}{4|{\bf p}|\sqrt{ \frac{\xi}{2}
\left(is-\frac{\lambda^{4}}{2}\xi p_{\mu}p^{\mu}\right)
\left(is+\frac{\lambda^{4}}{2}\xi {\bf p}^{2}\right)}} \right]
\right. \nonumber \\
\left.
+\mbox{erfi}\left[q\,\lambda^{2}|{\bf p}|\left(\frac{\xi}{2}\frac{is-\frac{\lambda^{4}}{2}\xi p_{\mu}p^{\mu}}
{is+\frac{\lambda^{4}}{2}\xi {\bf p}^{2}}\right)^{1/2}
\right. \right. \nonumber \\
\left. \left.
-\lambda^{2}\eta\frac{\left|\lambda^{4}\xi p_{\mu}\left(k_{1}+k_{4}\right)^{\mu 0}p_{0}p^{j}
-2p_{\mu}\left(k_{1}+k_{4}\right)^{\mu j}\left(is+\frac{\lambda^{4}}{2}\xi{\bf p}^{2}\right)\right|}{4|{\bf p}|\sqrt{ \frac{\xi}{2}
\left(is-\frac{\lambda^{4}}{2}\xi p_{\mu}p^{\mu}\right)
\left(is+\frac{\lambda^{4}}{2}\xi {\bf p}^{2}\right)}}  \right] \right\}
\; .
\end{eqnarray}
%
%
%
It is not difficult to see that the previous integral has a good behavior when $p_{\mu}p^{\mu}<0$,
while it is divergent if $p_{\mu}p^{\mu}>0$. After a tedious calculus, the radial integral can be manipulated in order to help us to use the
Handbook's integral to obtain the result
\begin{eqnarray}\label{Sigma1npIntdsk1k4}
\Sigma_{1(np)}(m^{2},p^{\mu};k_{1}+k_{4})= \frac{ig}{48\pi^{2}}
\int_{0}^{\infty} \frac{ds}{s^{2}} \left(\frac{\lambda^{2}}{\pi s} \right)^{6}  \left(1+i\frac{\lambda^{4}}{2s}\xi p_{\mu}p^{\mu}\right)^{-3/2}
\, \times
\nonumber \\
\times \,\, \frac{e^{-is\left(m^{2}-i\varepsilon\right)}}{\left(1+i4\lambda^{2}/s\right)^{3}} \, e^{-\frac{\lambda^{4}}{4}\xi\left(k_{1\mu\nu}+k_{4\mu\nu}\right)^{2}
-\frac{\lambda^{8}}{4}\xi^{2}\left[ p_{\mu}\left(k_{1}+k_{4}\right)^{\mu0} \right]^{2}\left(is+\frac{\lambda^{4}}{2}\xi {\bf p}^{2}\right)^{-1}}
\, \times
\nonumber \\
\times \,
e^{\frac{\lambda^{4}\xi}{2{\bf p}^{2}}\left|\frac{\lambda^{4}}{2} \xi \left(is+\frac{\lambda^{4}}{2}\xi{\bf p}^{2}\right)^{-1}
\! p_{\mu} \left(k_{1}+k_{4}\right)^{\mu0}
p_{0}p^{j}-p_{\mu}\left(k_{1}+k_{4}\right)^{\mu\,j}\right|^{2}\left(is+\frac{\lambda^{4}}{2}\xi {\bf p}^{2}\right)\left(is-\frac{\lambda^{4}}{2}\xi p_{\mu}p^{\mu}\right)^{-1}
} \, \times
\nonumber \\
\times \,
e^{\frac{\lambda^{4}\xi}{2{\bf p}^{2}is}\left|\frac{\lambda^{4}}{2}\xi \left(is+\frac{\lambda^{4}}{2}\xi{\bf p}^{2} \right)^{-1}\! p_{\mu}\left(k_{1}+k_{4}\right)^{\mu 0}p_{0}p^{j}-p_{\mu}
\left(k_{1}+k_{4}\right)^{\mu j} \right|^{2}\left(is+\frac{\lambda^{4}}{2}\xi{\bf p}^{2} \right)^{2}\left(is-\frac{\lambda^{4}}{2}\xi p_{\mu}p^{\mu}\right)^{-1}}
\; .
\end{eqnarray}
When the external $k$-momenta are null, that is, $k_{1}=k_{4}=0$,
it is direct to see that the result (\ref{Sigma1npIntdsk1k4})
reduces to the expression (\ref{Sigma1npsint}), with the factor $\left(1+i4\lambda^{2}/s\right)^{-3}$
due to the integrations $(k_{2},k_{3})$ in the loop.  Hence, we can write the integral in (\ref{Sigma1npIntdsk1k4}) as
\begin{eqnarray}\label{Sigma1npIntdsk1k4=0}
\Sigma_{1(np)}(m^{2},p^{\mu};k_{1}+k_{4}=0)= \frac{ig}{48\pi^{2}}
\int_{0}^{\infty} \frac{ds}{s^{2}} \left(\frac{\lambda^{2}}{\pi s} \right)^{6} \, \times
\nonumber \\
\times \, \left(1+i\frac{\lambda^{4}}{2s}\xi p_{\mu}p^{\mu}\right)^{-3/2}
\!\!\left(1+i4\lambda^{2}/s\right)^{-3}
e^{-is\left(m^{2}-i\varepsilon\right)}
\; .
\end{eqnarray}

\noindent Notice, that NCY introduces a huge difficulty concerning the solution of this integral.  We can see that the integral variable $s$ is present in every term of the integral.  So, there is mix of high power of $s$ together with square root of $s$ and more than one term with negative high power of $s$, which means high power of $s$ in the denominator.  And the numerator cannot be eliminated by the denominator because $s$ is inside non trivial expressions with the NC parameter.
Unfortunately, the solution of this integral requires numerical computation work, which is out of the scope of this paper.
%
%

%
%



\section{Conclusions and final remarks}
\renewcommand{\theequation}{5.\arabic{equation}}
\setcounter{equation}{0}

The quest to understand the Early Universe physics has many candidates and one of them is NCY which has its measure linked to the Planck scale.  This $\theta$-parameter, nowadays, has two features.  It can be constant, which causes the Lorentz invariance breaking, or as a coordinate of a NC phase-space that is formed by $(x, p, \theta^{\mu\nu}, k_{\mu\nu})$ where $k_{\mu\nu}$ is the canonical momentum conjugate to $\theta^{\mu\nu}$.  It can be demonstrated that $k_{\mu\nu}$ is also connected to the Lorentz invariance \cite{ammo}.

Considering $\theta^{\mu\nu}$ as a coordinate we have in the current literature, two formulations that are directly related to each other.  The standard DFR considers $\theta^{\mu\nu}$ as a coordinate and the so-called DFR-extended, i.e., DFR$^*$, which recognize the existence of $k_{\mu\nu}$.  As we saw in this work, we can construct an interactive QFT in this DFR$^*$.

Having said that, in this work we have analyzed the self-quartic interaction for a scalar field through Feynman diagrams formalism and the vertices calculation were accomplished.  After obtaining these Feynman rules, we computed the radiative corrections to one loop order for the propagator of the model.  The objective here is to analyze the mixing IR/UV divergences.

Concerning the radiative corrections we have investigated two cases: $k_{\mu\nu} =0$ and $k_{\mu\nu} \neq 0$.  The first one, although it is not an element of the DFR$^*$ phase-space, as we have explained, it is useful to compare with the $k_{\mu\nu} \neq 0$ case, namely, $k_{\mu\nu}=0$ is a toy model in DFR$^*$ phase-space.  Hence, both the self-energy of planar and non-planar diagrams presents difference concerning the NC parameter presence.  The non-planar diagram is infected with NCY.  The second case, $k_{\mu\nu} \neq 0$, presents both planar and non-planar scenarios contaminated with NCY.


\section{Acknowledgments}

\noindent EMCA would like to thank CNPq (Conselho Nacional de Desenvolvimento Cient\' ifico e Tecnol\'ogico), Brazilian scientific support agency, for partial financial support.


%
%

%

\end{document}